\documentclass[prb,reprint,amsmath,amssymb,superscriptaddress]{revtex4-1}
\usepackage[utf8]{inputenc}
\usepackage{graphicx}
\usepackage{hyperref}
\usepackage{physics}
\usepackage{color}
\usepackage{url}

\begin{document}

\preprint{APS/123-QED}

\title{Signatures of clean phases in many-body localized quantum circuits}
\author{Kaixiang Su}
\affiliation{Dept. of Physics, University of California at Santa Barbara, Santa Barbara, CA.}
\altaffiliation{Dept. of Physics, Peking University, Beijing, China}
\altaffiliation{Dept. of Physics, Cornell University, Ithaca, NY, USA}

\author{Michael J. Lawler}%
 \email{mjl276@cornell.edu}
\affiliation{Dept. of Physics, Cornell University, Ithaca, NY, USA.}
\altaffiliation{Dept. of Physics, Applied Physics, and Astronomy, Binghamton University, Binghamton, NY, USA.}

\date{December 2021}

\begin{abstract}
    Many-body phenomena far from equilibrium present challenges beyond reach by classical computational resources. Digital quantum computers provide a possible way forward but noise limits their use in the near-term. We propose a scheme to simulate and characterize many-body Floquet systems hosting a rich variety of phases that operates with a shallow depth circuit. 
    Starting from a ``clean" periodic circuit that simulates the dynamical evolution of a Floquet system,
    we introduce quasi-periodicity to the circuit parameters to prevent thermalization by introducing many-body localization.  By inspecting the time averaged properties of the many-body integrals of motion, the phase structure can then be probed using random measurements. This approach avoids the need to compute the ground state and operates at finite energy density. We numerically demonstrate this scheme with a simulation of the Floquet Ising model of time-crystals and present results clearly distinguishing different Floquet phases in the absence of quasi-periodicity in the circuit parameters. Our results pave the way for mapping phase diagrams of exotic systems on near-term quantum devices.
\end{abstract}

\maketitle
\section{Introduction}
Numerical analysis of certain condensed matter problems constitute an important theme in condensed matter theory. However it is well known that certain problems exceed the computational power of classical computers. A faithful calculation of the properties of certain states of matter requires exponential computational resources, which inevitably fails when we try to increase the system size. It is then no wonder that condensed matter physicists become excited about the advances made in quantum computation and the promise that quantum computers could provide a way to overcome these computational obstacles.

Some of the attempts to utilize quantum computational powers include a variational determination of the groundstates of certain Hamiltonians. This direction is termed variational quantum eigensolver (VQE) and some progress has already been made in this area\cite{vqe2014,vqe2017}. Although very useful for small sized systems, such schemes face difficulties when increasing system sizes because it has been proven that many classes of quantum Hamiltonians are QMA-complete\cite{qma_review}. Among these QMA-complete Hamiltonians are the Bose-Hubbard models\cite{qma_bosehubbard}, which are of great interest to the condensed matter community. As a result, it is unclear whether a protocol utilizing VQE to study condensed matter systems beyond the reach of classical computers is attainable. 

\begin{figure}
    \centering
    \includegraphics[width=\columnwidth]{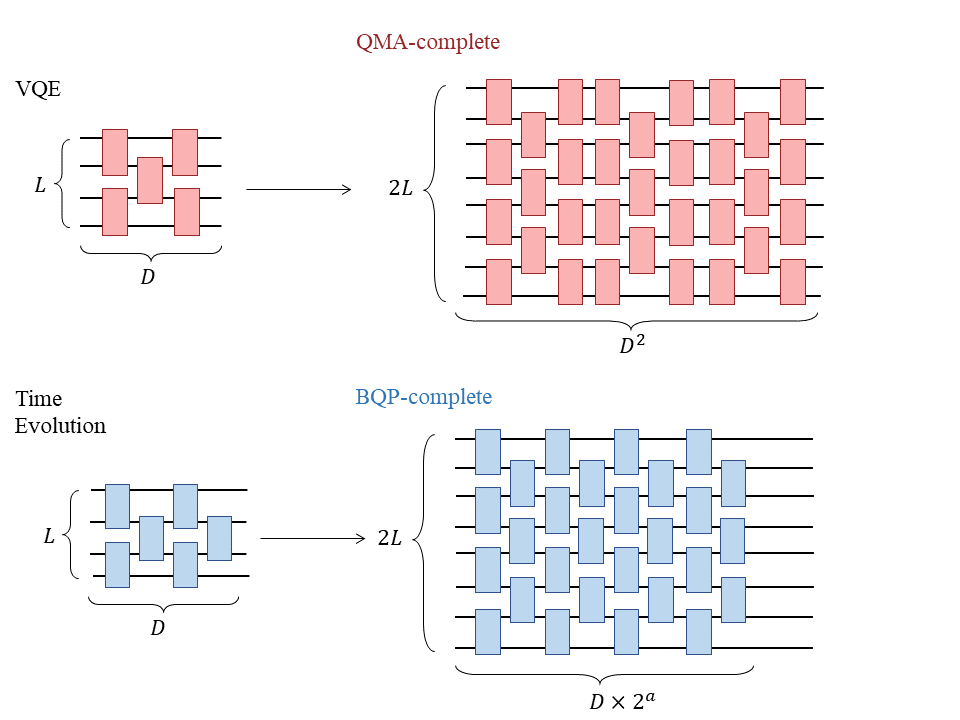}
    \caption{Differences between the two classes of operators. QMA-complete algorithms such as VQE might be suitable for small system sizes, but become inefficient when system size grows large. BQP-complete problems such as Hamiltonian simulation are efficient with system size scaling (polynomial for example). }
    \label{fig:QMAvsBQP}
\end{figure}

Contrary to the fact that VQE belongs to QMA-complete problems, simulating local Hamiltonian evolution on a quantum computer is known to be BQP-complete\cite{Bauer2020,Kitaev}. An illustration of the differences between these two class of algorithms are illustrated in fig. \ref{fig:QMAvsBQP}. Although this does not guarantee an efficient algorithm to study the many body problems, we do get some motivation from this line of thinking, which leads us to consider whether there exists an algorithm utilizing dynamical simulation on quantum computers. From an experimental perspective, compared to a thorough determination of the eigenstates, it seems more feasible to compare data with dynamical simulations of quantum systems obtained from quantum computers. With the advances made in constructing quantum computing devices and platforms, some of the near-term quantum devices already provide a chance to approach the limit of classically computational powers of certain problems. Most notably are those using trapped ions\cite{Smith2019, smith2020crossing}, superconducting qubits\cite{google_timecrystal, CASqubits} and other experimental platforms such as optical lattices and cold atoms. Quite a few simulation and detection schemes have been proposed to construct and study certain interesting many body phases of matter. \cite{Smith2019, smith2020crossing}

In recent years a certain type of many body systems has attracted significant interest: periodically driven Floquet systems. They are intersting because in many cases they can exhibit certain properties, topological ones for example\cite{topofloquet}, not previously seen in equilibrium systems.  One class of Floquet systems of particular interest might be the recently proposed time crystals. In this paper we identify a scheme to simulate Floquet quantum many body phases using a noisy intermediate scale quantum computer (IBM-Q for example). This scheme avoids thermalization of the system caused by time evolving finite energy states and identifies ways to detect the phase using measurement outcomes from a quantum computation procedure. We support this scheme with numerical results demonstrating its validity in a model of time crystals. 

This paper is organized as follows: In section 2 we present the ideas underpinning the proposed scheme and explain how we avoid thermalization through many-body localization. In section 3 we present the quantum circuits that simulate the Floquet phases. In section 4 we present measurement protocols that enable the identification of phases.  Some of our numerical results to demonstrate the validity of our approach is presented in section 5. We conclude by identifying interesting future directions including potential alternative schemes related to measurement induced dynamical phase transitions. 

\begin{figure}
    \centering
    \includegraphics[width=\columnwidth]{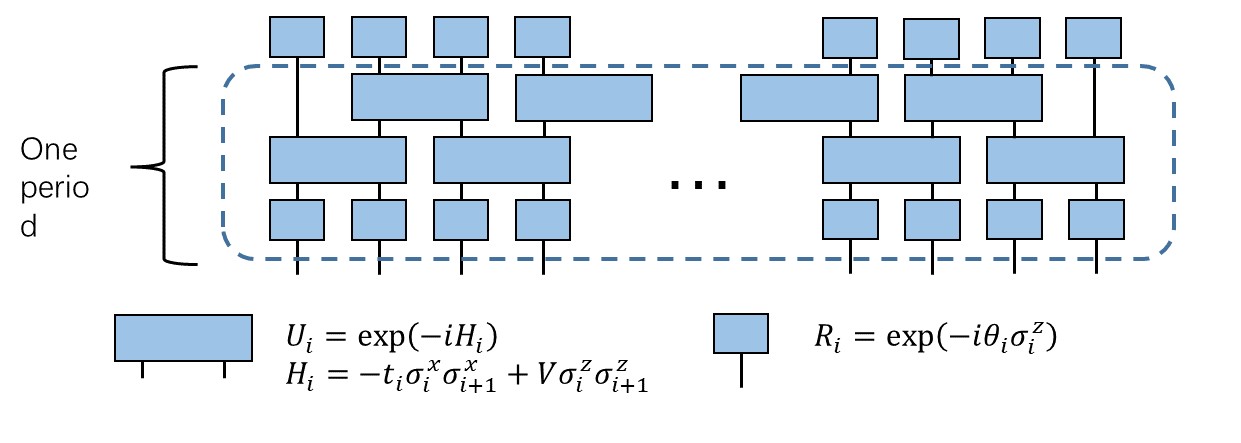}
    \caption{An illustration of the circuit setup. The whole evolution consists of identical groups of layers. Each group, or period, is made up of three layers of quantum gates. The first layer is composed of single qubit phase rotations, and the second and third layers are composed of two qubit gates that simulate the two-body interactions.}
    \label{fig:circuit}
\end{figure}

\section{Avoiding Thermalization Through Localization}
One would think to start from a clean system and study its dynamics on a quantum computer, since these systems are the simplest in some sense. However when it comes to simulating Floquet dynamics of a clean system there comes a severe issue one must resolve to get sensible results---the issue of thermalization. Since we are seeking a general algorithm, the systems we deal with then necessarily have generic interactions, and are not integrable. Such systems are believed to thermalize after a sufficient amount of evolution time has passed. Many subtle quantum phases do not persist to finite temperatures, so by only looking at the long time results of evolution, we lose signatures of the phase diagrams.

\begin{figure}
    \centering
    \includegraphics[width=\columnwidth]{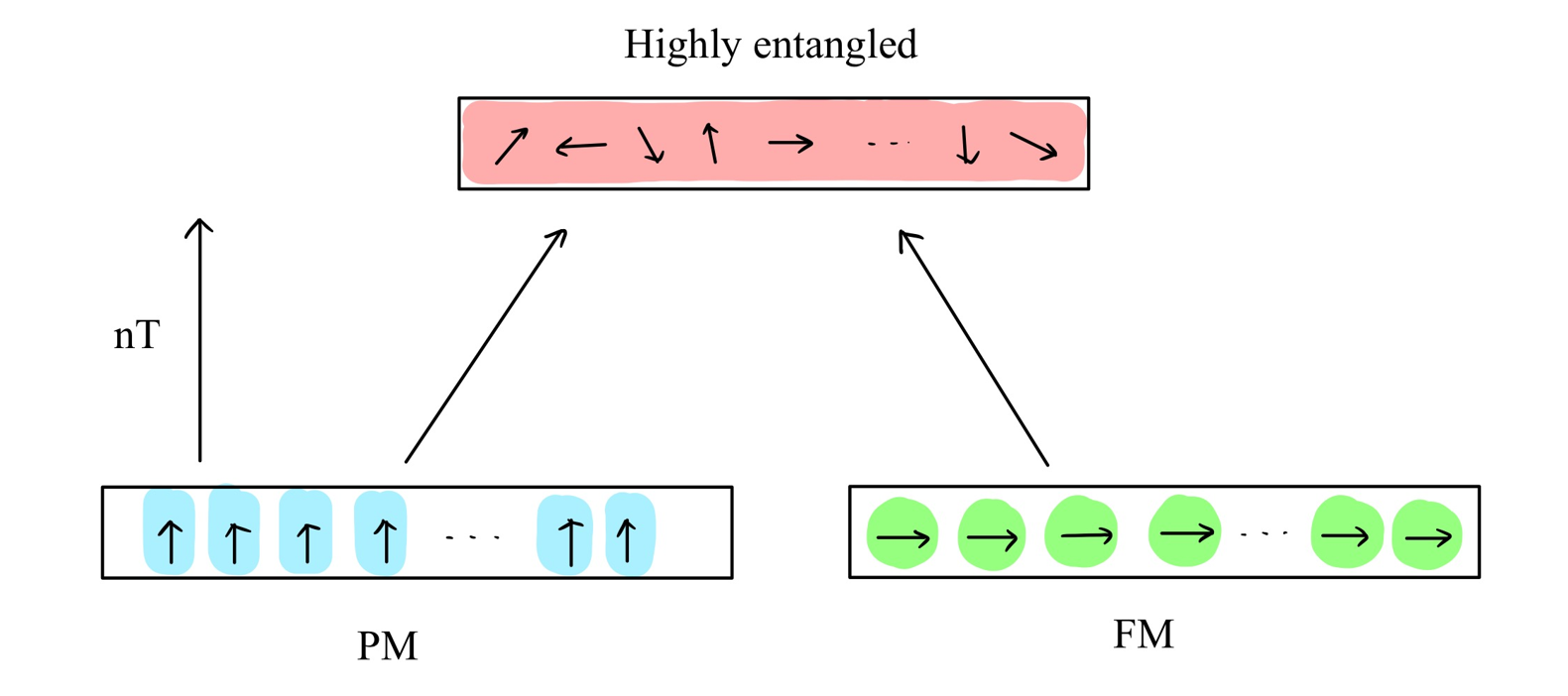}
    \caption{Generic non-integrable systems are expected to thermalize after long time evolution. For Floquet systems the effective final temperature is infinity. Starting from a product state representing two distinct phases, eventually the evolution will take the product states to highly entangled state in which no signatures of the original phases can be traced out. }
    \label{fig:thermal}
\end{figure}

A way to 'protect' these quantum phases from thermalization is proposed in \cite{ashvin} utilizing many-body localization. Many-body localization is a generalization of the Anderson localization\cite{mbl_rmp}. When we induce certain degrees of disorder into the system, the disjoint parts of the system will only interact with each other weakly. Thus the local information is largely preserved during the evolution. There are many ways to devise this disorder, for example by random variables or more recently proposed 'stark localization'\cite{starkmbl}. In our model design we use the quasi-periodicity setting merely for simplicity. That is, we let our parameter $t$ and $\theta$ have a periodicity not commensurate with the lattice. More specifically the parameters have the following fashion:
    \begin{align}
        & t = t_0 + t_1 \cos (2\pi k i +\phi), \\
        & k = \frac{\sqrt{5}-1}{2}
    \end{align}
where i is the position index of the gate. Remarkably, the many-body localized systems have an extensive set of local integrals of motion (LIOMs). Usually they are a dressed version of some local products of Pauli operators denoted by $\tau_i$ as compared to the original Pauli operators $\sigma_i$. We will see below that these Integrals of motion also play a crucial role in our detection of specific phases of matter. 
    
There is, however, no guarantee that the system after adding disorder shares the same phase structure with the original system. Thus, only those phases that persist when we induce localization will be detected with our approach. The specific model system we use, on the other hand, has this desired feature that the disordered system has the same phase diagram as the clean one. A study of the generic validity of this feature is interesting in its own right. 

\section{Model}
To illustrate the applicability of our scheme we design a Floquet quantum circuit to see if its phase structures can be detected. A Floquet system, by its name, is just a periodically driven system. The periodicity can easily be incorporated into circuits arrangements. We can simply construct a group of quantum gates of several layers. By repeatedly applying this group of quantum gates we are then mimicing a periodic evolution. More specifically our model is a circuit analogy of the Floquet Ising model\cite{Khemani1}. An illustration of the scheme is shown in fig. \ref{fig:circuit}.
The model we design is a variation of the famous Floquet Ising model first discussed in\cite{Khemani1}. Originally proposed as a Floquet topological phases of matter, it has a rich phase diagram containing phases not present in its equilibrium cousins. Notably, other than the paramagnetic(PM) and Ferromagnetic(FM) phases, it hosts also the time crystal phase (also called the $\pi$ spin glass phase) and the other $0\pi PM$ phase. We should mention that although this model is interesting on its own, it mainly serves as a demonstration of our methods here. The scheme we used can actually be applied to more general models and settings. \\
\begin{figure}
    \centering
    \includegraphics[width=\columnwidth]{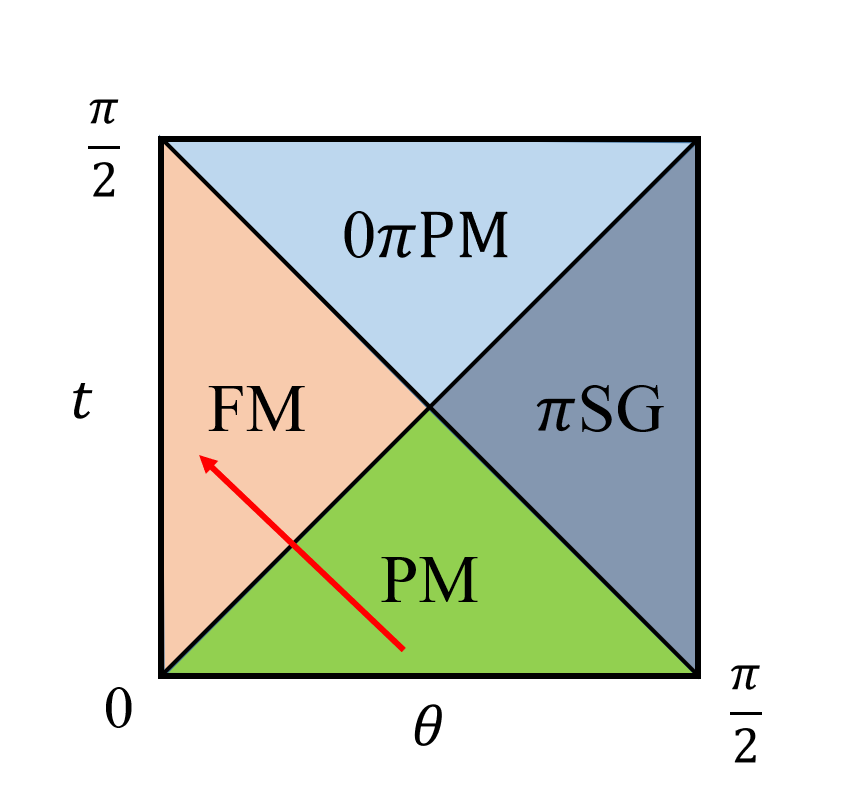}
    \caption{Phase diagram of the original model\cite{Khemani1}. The two axis are phase rotation angle and hopping strength respectively. The phase diagram is symmetric around $\theta = \pi/2$ and $t = \pi/2$. The red arrow depicts our scanning scheme detailed in the results section. }
    \label{fig:phase}
\end{figure}
The circuits we design to mimic the Floquet evolution consists of three layers. The first layer is a single phase rotation on each qubits. The second and third layers are two qubit gates acting on nearest neighbors. The phase rotation is analogous to the transverse field, while the two qubit gates incorporate both the magnetic nearest neighbor interactions and other interactions that make the system a non-integrable many body system. The exact phase boundaries are certainly different from the original phase diagram. But in a sense to be explained below, the phase structure of our circuits should be similar to the phase diagram (fig. \ref{fig:phase}).  

\section{Measurement Protocols}
Now even if we realized a stable phase of matter in our Floquet dynamics simulation of an Ising-like model, what are the probes or measurements we could use to discuss properties of its phases? The notion of an order parameter certainly exists in equilibrium states of matter, but whether the same logic could be applied to Floquet simulations is not so clear. In the following we describe methods we use to resolve these problems. 

\subsection{Order Parameter Dynamics}
For parameters falling in different phase regimes, their LIOMs often take different forms. For example, in the paramagnetic phase the LIOMs are dressed versions of the Pauli-Z operators while in the ferromagnetic phase the LIOMs are dressed versions of the Pauli-XX operators. It would seem that the LOIMs are related to order parameters.

However, knowing a relation to order parameters does not readily give us a way to tell, after a simulation, what LIOMs do we actually have in the experiment. To accomplish a detection, we utilize the fact that the LIOMs are invariant upon time evolution. If we pick some operator sufficiently close to the LIOMs, then most of its information gets preserved during the evolution. By this we simply mean, that the original operators can be decomposed into a main operator and a few other operators that have relatively smaller sizes. For example, when we are in the paramagnetic phase, most of the information of the Pauli-Z operators will be preserved while most of the information of the Pauli-X operators are lost. So if we start with a Pauli-X operator, in the Heisenberg representation the operator always changes rapidly. By doing a proper time averaging, those fast changing parts vanish while those constant parts remain. 
    
\begin{figure}
    \centering
    \includegraphics[width=\columnwidth]{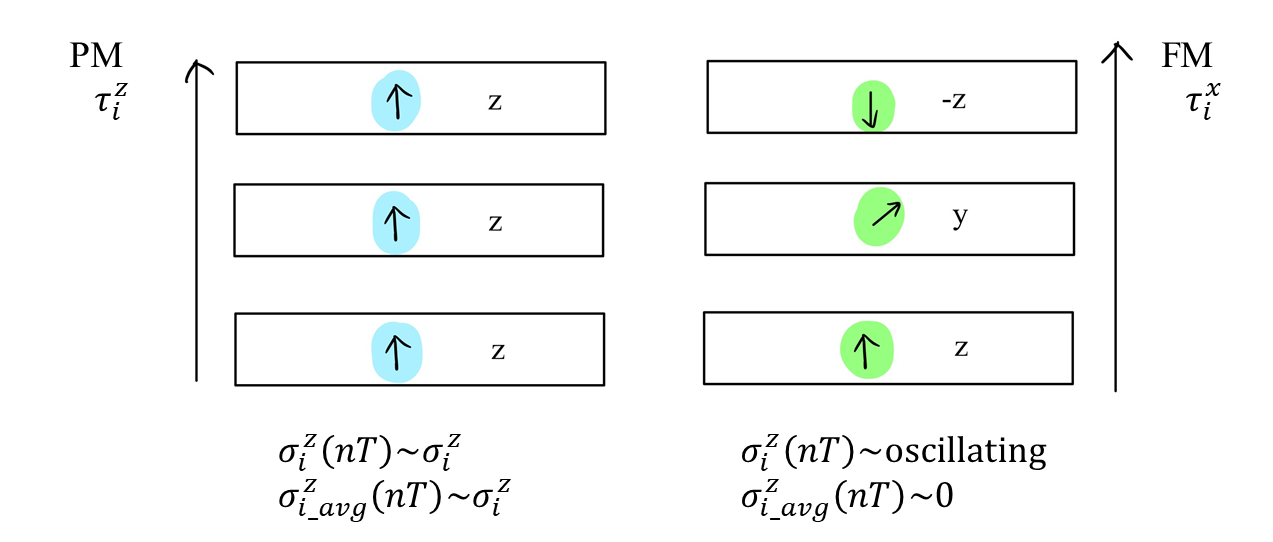}
    \caption{Effects of LIOMs (Local Integrals of Motion) on the evolution of physical observables. In Heisenberg picture those LIOMs preserve their initial values while operators other than the LIOMs oscillate rapidly. Under time average only information of the LIOMs are preserved. }
    \label{fig:DOP1}
    \end{figure}

We can present a more precise notion of this. We can define the following quantity to quantify how close is the evolution of order parameter to being locally conserved 
    \begin{equation}
        \big|\sigma_i^x \sigma_j^x(nT)\big|  = {\mathrm Tr} [(\sigma_{i}^x(nT) \sigma_{j}^x (nT))_{avg}^2]/2^N
    \end{equation}
with the time average of an operator defined by
    \begin{equation}
        \hat{O}_{avg}(nT) = \frac{1}{n}\sum_{j = 1}^n \hat{O}(jT)
    \end{equation}
In other words, it is the size of the time averaged normal order parameter. The operators are in the Heisenberg representation, and being averaged with respect to different time steps. Physically this measures how close is the operator under consideration to the real local integrals of motion. Since the real local integrals of motion are invariant under time evolution, the time average just gives back the original operators. On the other hand, the parts different from LIOMs are changing rapidly and vanish upon being averaged. An illustration of this is presented in fig. \ref{fig:DOP1}.

We can estimate how the circuit depth scales with the system size. All we need is to evolve enough time steps so that the components in the original order parameter get 'smeared out' under time averages. Since the order parameters are quite local, the time scale associated with this is equivalent to the relaxation time, which only depends on the support range of the operator in a power-law fashion\cite{LIOMs,LIOM_review}. Since the support range of the operator is roughly fixed with increasing size for a given parameter point in the phase diagram, we estimate that the circuit depth does not scale with system size. However, since close to the critical point the support range of the local integrals of motion will tend to become the size of the system, we expect the depth of the circuit to increase siginificantly close to the critical point. 
    
\subsection{Random Measurements}
The size of an operator is not an easy one to obtain with traditional qubits measurements. Here we employ a method inspired by \cite{randscramb} to take advantage of the so called random measurements. We note here that the full derivation of the formulas used here and a detailed description of the setup can be found in the original paper. Here the only difference is that we will be interested in time-averaged quantities, and this extra subtlety is resolved at the end of this subsection. 

\begin{figure}
    \centering
    \includegraphics[width=\columnwidth]{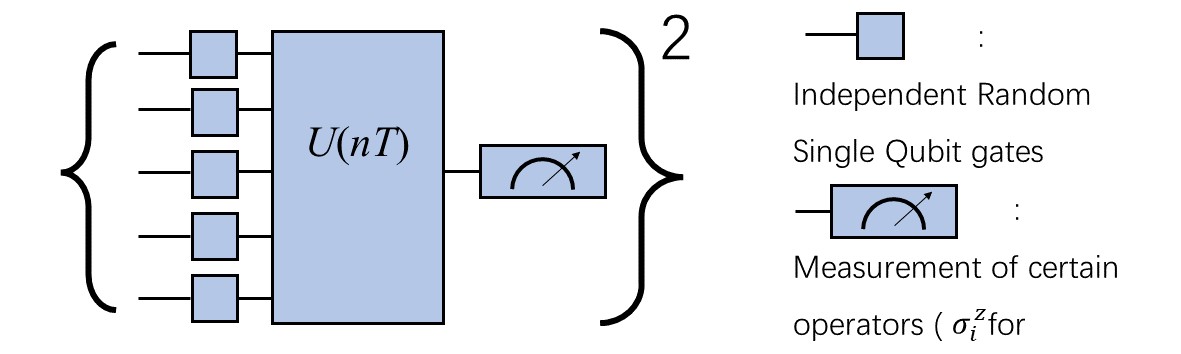}
    \caption{The setup of random measurements. The initial state is drawn from an ensemble which is independently Haar random on each spin. A Floquet circuit is then applied to this random state and normal measurements are taken. By averaging the square of the measurement outcomes with respect to the random state ensemble, for example, one can get information regarding the size of the measurement operator in the Heisenberg picture. }
    \label{fig:DOP2}
    \end{figure}

Generally there are two random measurement protocols that can achieve such measurements, one based on global random states and the other based on locally random states. Since the generation of global random states with quantum circuits requires comparatively more layers and is yet another independent direction in quantum computations, we will restrict ourselves to the protocol based on local random states. Basically it goes like this: we prepare a product state but each qubit is randomized independently. This state then undergoes a time evolution after which the measurement of the desired operator is taken. The size of the operator can be extracted from the statistical correlations of these different results when different initial states are prepared (the statistical correlations have nothing to do with probabilistic outcomes due to uncertainty principles). An illustration is presented in fig. \ref{fig:DOP2}. 

More specifically, we prepare a product state in the computational basis with all qubits set to '0'. We will denote this state as $\ket{k_0}$. We then form an ensemble of states by flipping a total of $m$ qubits, which consists of $2^m$ different states. Sampling now the local random gates and denote this specific instance of random circuits as $u$, we then calculate the quantity
    \begin{equation}
        \langle \hat{O}(nT) \rangle_u \langle \hat{O}(nT) \rangle_u
    \end{equation}
We now need to do a weighted sum over all the $2^m$ states in the ensemble, with weights $(-1/2)^{\textrm{\# of flipped qubits}}$. Then we sample other local random circuits and repeat this process. Eventually this gives the equality
    \begin{align}\label{eq:Oexpect}
    \begin{split}
        \overline{\langle \hat{O} \rangle_u \langle \hat{O} \rangle_u}  = &\frac{1}{3^L}(\frac{3}{4})^n(\frac{1}{2})^{L-n} \textrm{Tr} (\prod_{j\leq n} \textrm{Swap}_j(\hat{O}_j \otimes \hat{O}_j) \\
        &\prod_{k>n}(\hat{O}_k\otimes\hat{O}_k + \textrm{Swap}_k (\hat{O}_k \otimes \hat{O}_k))). 
    \end{split}
    \end{align}
It important to note that the general operator $\hat{O}$ is represented in a matrix product form. Different $\hat{O}_j$s should be understood as having two extra legs contracted with $\hat{O}_{j-1}$ and $\hat{O}_{j+1}$. The swap operator, after acting on the tensor product and being traced over, gives contraction of $\hat{O}_k$ and $\hat{O}_k$.

The rather complicated expression of Eq. \ref{eq:Oexpect} will look very simple and ideal for our case if we take $m$ to be $L$, in which case the right hand side of the equation is just proportional to $\textrm{Tr} (\hat{O}^2)$. So one can view the $m$ parameter as the level of approximation. However, the quantity we aim to include in the expression is $\hat{O}_{avg}(nT)$, which according to the properties of LIOMs should be a local operator. Thus we should need a relatively small $m$ to still give good enough results, in other words the results converge rapidly with increasing $m$. 
    
Some may wonder if this is readily applicable to our scheme since what appears in the dynamics of order parameters is not a genuine operator but a time average of it. In fact, applying to our scheme require one to evaluate
    \begin{equation}
        \langle \hat{O}_{avg}(nT) \rangle = \frac{1}{n} \sum_{j = 1}^n \langle \hat{O}(jT) \rangle
    \end{equation}
A convenient way would be to record the measurement results for each time step. One then needs to square the time averaged measurement results to proceed with what we described above. We should note that in our classical simulations we do not take into account finite $m$ effects and the statistical error of sampling random unitaries. Classical simulation is able to directly compute the size of the time-averaged order parameters since we study a small system.
    
\section{Results}
We now present our results. Our first result is the order parameter dynamics for different parameters. More specifically, we choose two points in the parameter space, one with $t = 0.2, \theta = 0.8$ (PM phase) and another with $t = 0.8, \theta = 0.2$ (FM phase) and see how our the dynamics of the order parameter behaves for these two circuits individually. We can see that for the Pauli-XX order parameter, the value of the order parameter dropped close to zero with finite time steps in the paramagnetic phase, while remaining finite throughout the time steps we take in the ferromagnetic phase. This is consistent with the intuition that FM phase has a $XX$-like order.

\begin{figure}
    \centering
    \includegraphics[width=\columnwidth]{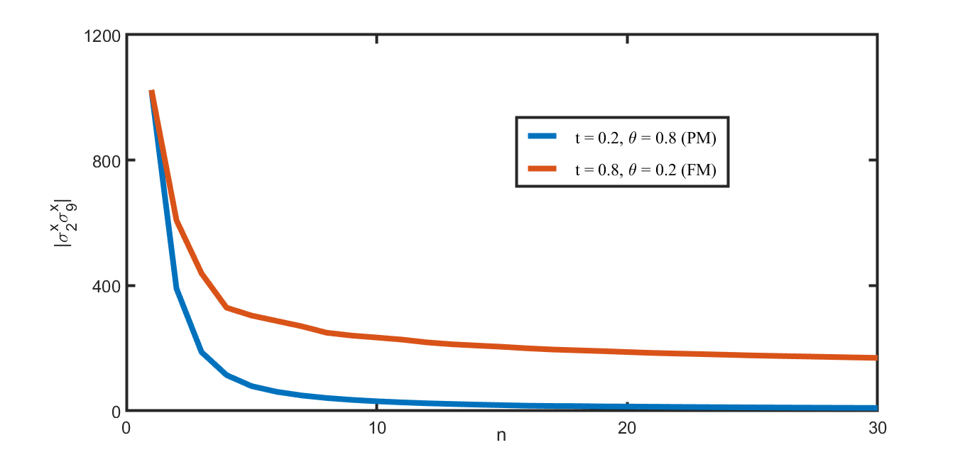}
    \caption{Behaviours of the order parameters dynamics in two different phases. The order parameter is chosen as $\sigma_i^x \sigma_j^x$ which is the order parameter for the FM phase. In the PM phase this order parameter dropped to zero in a finite number of steps. However in the FM phase the order parameter stays finite within the time steps taken. }
    \label{fig:results1}
    \end{figure}
    
We then obtain results by scanning through a range of parameter values that cross the phase boundary. Ideally one could wait long enough to see the 'true' saturation value of the order parameters. In our simulation we treat 1000 time steps more than as enough to produce reliable results. In this case one can see that the saturated value of order parameter starts to deviate from 0 at certain values of $t$. In simulations on NISQ devices, one certainly has no access to this amount of time steps due to noise limitations that restrict the circuit's gate depth. So we also plot results we obtain by restricting to 30 time steps, a fair amount for a near-term quantum computer. By utilizing the previous time steps, we extend beyond 30 time steps by fitting them to a power-law decay. This heuristic approach seems a stronger indicator of our known long time results than just by looking at the 30 time steps alone. So either with 1000 time steps or 30 time steps we find signatures of the phases of matter in our time-and-ensemble averaged order parameters. 

\begin{figure}
    \centering
    \includegraphics[width=\columnwidth]{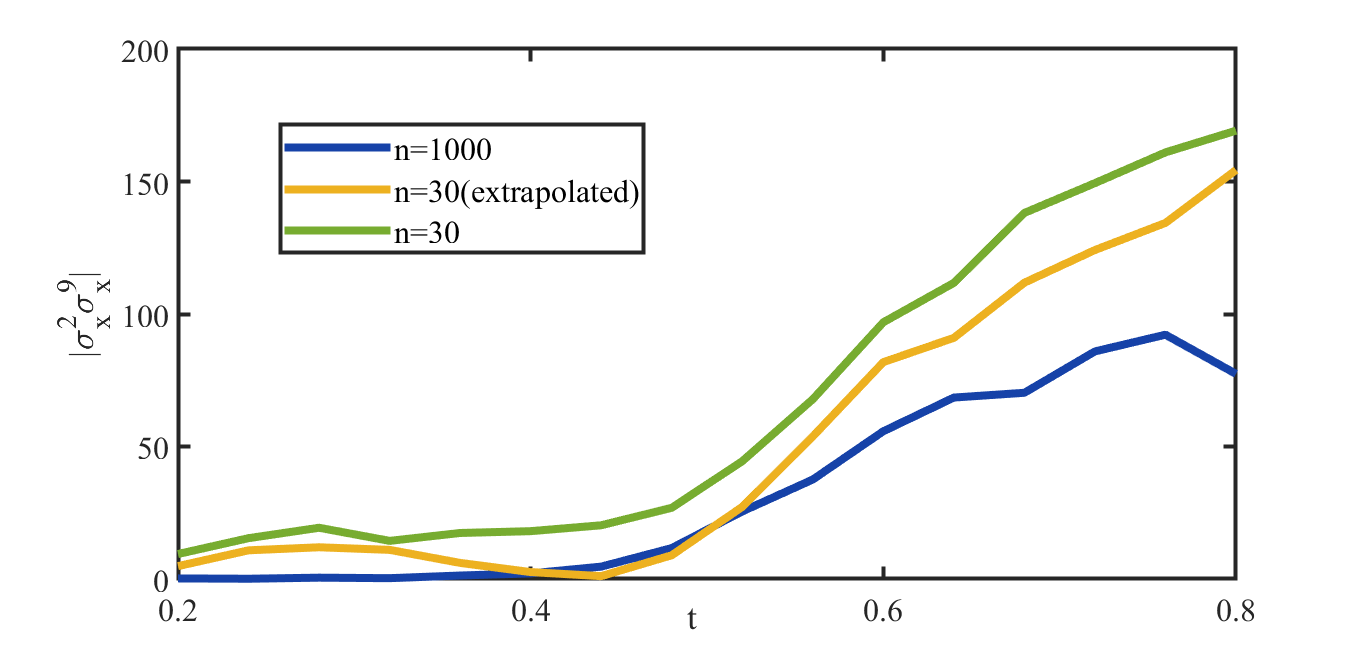}
    \caption{A simulation of scanning through the phase boundaries. An illustration of the scanning trajectory is depicted as the red arrow in fig. \ref{fig:phase}. The extrapolated results for 30 time steps agree qualitatively with the ideal case (n = 1000).  }
    \label{fig:results2}
    \end{figure}
    
\section{Conclusion \& Outlooks}
To summarize, in this paper we proposed a scheme to stabilize and dynamically detect Floquet phases of matter on a near term quantum computer. The central concept we seek to utilize is many-body localization, which prevents our circuits to thermalize and render a featureless phase diagram. Instead, with many-body localization the circuits in different phases show distinct behaviours even after undergoing significantly long times. The distinction between different phases is encoded in the LIOMs of the circuits, which is detected by measuring time-and-ensemble averaged order parameters. 

A crude estimate for the quantum volume (the number of qubits times the gate depth accessible to a quantum information processor) required to exceed the classical simulation power of this problem for a quantum computer would be approximately 30*20. We expect this scheme could be deployed on the IBM Q quantum computing platform upon adaptation in the near term, especially on their recently announced 127-quantum bit (qubit) 'Eagle'\cite{eagle}. 

Some future directions to look into would be to include noise expected in a quantum computer into our scheme and assess how the noise changes the validity of our proposal. Specifically how does noise alter the behaviours of LIOMs in the presence of localization? It is also worth noting that the random measurement schemes proposed in this paper seem natural fits to explore more about the local integrals of motion in many-body localized systems, which are crucial for understanding many-body localized phases. Also, in our approach to study the phase diagram of a model, we effectively take the clean model and insert randomizing gates that randomize the parameters. An alternative approach could be to insert random measurements and drive the clean model through a measurement induced phase transition to an area law entanglement phase\cite{fisher,nahum,nandkishore}. At some level, this alternative approach is a space-time rotation of the approach discussed in this manuscript\cite{Grover} and may similarly allow the computation of phase diagrams.

\acknowledgements
This material is based upon work supported by the National Science Foundation under Grant No. OAC-1940260.

\bibliographystyle{unsrtnat}
\bibliography{reference.bib}

\end{document}